\documentclass[journal=ancac3,manuscript=article,layout=twocolumn]{achemso}
\setkeys{acs}{articletitle=true,etalmode=truncate,maxauthors=20}

\usepackage[version=3]{mhchem} % Formula subscripts using \ce{}
\usepackage[T1]{fontenc}       % Use modern font encodings
\usepackage{graphicx}
\usepackage{color}
\usepackage{float}

\author{Sudipta Dubey}
\affiliation[Univ. Grenoble Alpes]
{Univ. Grenoble Alpes, CNRS, Institut N\'{e}el, 38000 Grenoble, France}
\altaffiliation
{These authors contributed equally to this work}

\author{Simone Lisi}
\affiliation[Univ. Grenoble Alpes]
{Univ. Grenoble Alpes, CNRS, Institut N\'{e}el, 38000 Grenoble, France}
\altaffiliation
{These authors contributed equally to this work}

\author{Goutham Nayak}
\affiliation[Univ. Grenoble Alpes]
{Univ. Grenoble Alpes, CNRS, Institut N\'{e}el, 38000 Grenoble, France}
\altaffiliation
{These authors contributed equally to this work}

\author{Felix Herziger}
\affiliation[Univ. Grenoble Alpes]
{Univ. Grenoble Alpes, CNRS, Institut N\'{e}el, 38000 Grenoble, France}

\author{Van-Dung Nguyen}
\affiliation[Univ. Grenoble Alpes]
{Univ. Grenoble Alpes, CNRS, Institut N\'{e}el, 38000 Grenoble, France}

\author{Toai Le Quang}
\affiliation[Univ. Grenoble Alpes]
{Univ. Grenoble Alpes, CNRS, Institut N\'{e}el, 38000 Grenoble, France}

\author{Vladimir Cherkez}
\affiliation[Univ. Grenoble Alpes]
{Univ. Grenoble Alpes, CNRS, Institut N\'{e}el, 38000 Grenoble, France}

\author{C\'{e}sar Gonz\'{a}lez}
\affiliation[CEA SPEC]
{SPEC, CEA, CNRS, Universit\'{e} Paris-Saclay, CEA Saclay, 91191 Gif-sur-Yvette Cedex, France}
\altaffiliation
{Departamento de F\'{i}sica Te\'{o}rica de la Materia Condensada and Condensed Matter Physics Center (IFIMAC), Facultad de Ciencias. Universidad Autonoma de Madrid, E-28049 Madrid, Spain}

\author{Yannick J. Dappe}
\affiliation[CEA SPEC]
{SPEC, CEA, CNRS, Universit\'{e} Paris-Saclay, CEA Saclay 91191 Gif-sur-Yvette Cedex, France}

\author{Kenji Watanabe}
\affiliation[NIMS]
{National Institute for Materials Science, Tsukuba, 305-0044, Japan}

\author{Takashi Taniguchi}
\affiliation[NIMS]
{National Institute for Materials Science, Tsukuba, 305-0044, Japan}

\author{Laurence Magaud}
\affiliation[Univ. Grenoble Alpes]
{Univ. Grenoble Alpes, CNRS, Institut N\'{e}el, 38000 Grenoble, France}

\author{Pierre Mallet}
\affiliation[Univ. Grenoble Alpes]
{Univ. Grenoble Alpes, CNRS, Institut N\'{e}el, 38000 Grenoble, France}

\author{Jean-Yves Veuillen}
\affiliation[Univ. Grenoble Alpes]
{Univ. Grenoble Alpes, CNRS, Institut N\'{e}el, 38000 Grenoble, France}

\author{Raul Arenal}
\affiliation[LMA]
{Laboratorio de Microscop\'{i}as Avanzadas, Instituto de Nanociencia de Arag\'{o}n, Universidad de Zaragoza, 50018 Zaragoza, Spain}
\altaffiliation
{ARAID Foundation, 50018 Zaragoza, Spain}

\author{La\"{e}titia Marty}
\affiliation[Univ. Grenoble Alpes]
{Univ. Grenoble Alpes, CNRS, Institut N\'{e}el, 38000 Grenoble, France}

\author{Julien Renard}
\affiliation[Univ. Grenoble Alpes]
{Univ. Grenoble Alpes, CNRS, Institut N\'{e}el, 38000 Grenoble, France}

\author{Nedjma Bendiab}
\affiliation[Univ. Grenoble Alpes]
{Univ. Grenoble Alpes, CNRS, Institut N\'{e}el, 38000 Grenoble, France}

\author{Johann Coraux}
\affiliation[Univ. Grenoble Alpes]
{Univ. Grenoble Alpes, CNRS, Institut N\'{e}el, 38000 Grenoble, France}
\email{johann.coraux@neel.cnrs.fr}

\author{Vincent Bouchiat}
\affiliation[Univ. Grenoble Alpes]
{Univ. Grenoble Alpes, CNRS, Institut N\'{e}el, 38000 Grenoble, France}

\title[Defect-driven properties]
  {Weakly Trapped, Charged, and Free Excitons in Single-Layer MoS$_2$ in Presence of Defects, Strain, and Charged Impurities}


\begin{document}

\begin{abstract}
Few- and single-layer MoS$_2$ host substantial densities of defects. They are thought to influence the doping level, the crystal structure, and the binding of electron-hole pairs. We disentangle the concomitant spectroscopic expression of all three effects, and identify to which extent they are intrinsic to the material or extrinsic to it, \textit{i.e.} related to its local environment. We do so by using different sources of MoS$_2$ --- a natural one and one prepared at high pressure and high temperature --- and different substrates bringing varying amounts of charged impurities, and by separating the contributions of internal strain and doping in Raman spectra. Photoluminescence unveils various optically-active excitonic complexes. We discover a defect-bound state having a low binding energy of 20~meV, that does not appear sensitive to strain and doping, unlike charged excitons. Conversely, the defect does not significantly dope or strain MoS$_2$. Scanning tunneling microscopy and density functional theory simulations point to substitutional atoms, presumably individual nitrogen atoms at the sulfur site. Our work shows the way to a systematic understanding of the effect of external and internal fields on the optical properties of two-dimensional materials.
\end{abstract}

\section*{Main text}

Single-layer molybdenum disulphide (MoS$_2$) is a widely-studied candidate for future optoelectronics, where energy conversion is achieved with much lesser amounts of matter than with traditional three-dimensional materials, and a wealth of functionalities emerge from flexibility and transparency.\cite{Wang} The direct band-gap in the electronic band structure\cite{Li} is the key to light emission\cite{Mak,Splendiani} and conversion\cite{Sundaram} in single-layer MoS$_2$. Due to the reduced dimensionality, Coulomb interactions play a key role in this system and lead to a large exciton binding energy. More generally this system is a playground for testing many-body Coulomb interaction theories that should be able to describe excitonic complexes.\cite{Ugeda,Kidd,Efimkin} The current understanding is that the excitons in monolayer transition metal dichalcogenides have fast radiative lifetimes\cite{Poellmann,Wang_b,Jakubczyk} (sub-picosecond) and that the interlayer excitons in type-II junctions are long lived\cite{Rivera,Palummo} (in the nanosecond range at low temperature). Such excitons could be used for light-emission or light-harvesting devices, respectively. In both cases the presence of defects will induce non-radiative decay, reducing device efficiency, and/or modify the exciton emission energy. A chemical treatment eliminating defects allowed to demonstrate a photoluminescence quantum yield close to unity, and long lifetimes.\cite{Amani} Nevertheless the nature of defects in non-treated samples and their role in radiative and non-radiative recombination remain as open questions. To bring clearcut answers to this pressing question, well-characterised defects need to be investigated and their influence on the physical properties need to be understood. They could be intrinsic to the single-layer, \textit{e.g.} sulfur vacancies and substitutional atoms,\cite{Dolui,Kim,Tongay,Qiu,Lu,McDonnell,Addou,Noh,Komsa,Gonzalez} or extrinsic to it, in the form of charged impurities, either trapped in the MoS$_2$ substrate\cite{Lu} or adsorbed on it.\cite{Late,Li_b,Lembke,Jariwala} The latter is thought to limit the electronic mobility of MoS$_2$-based transistors below the phonon-limited value.\cite{Kaasbjerg}

\begin{figure*}[hbt]
 \begin{center}
 \includegraphics[width=142.9mm]{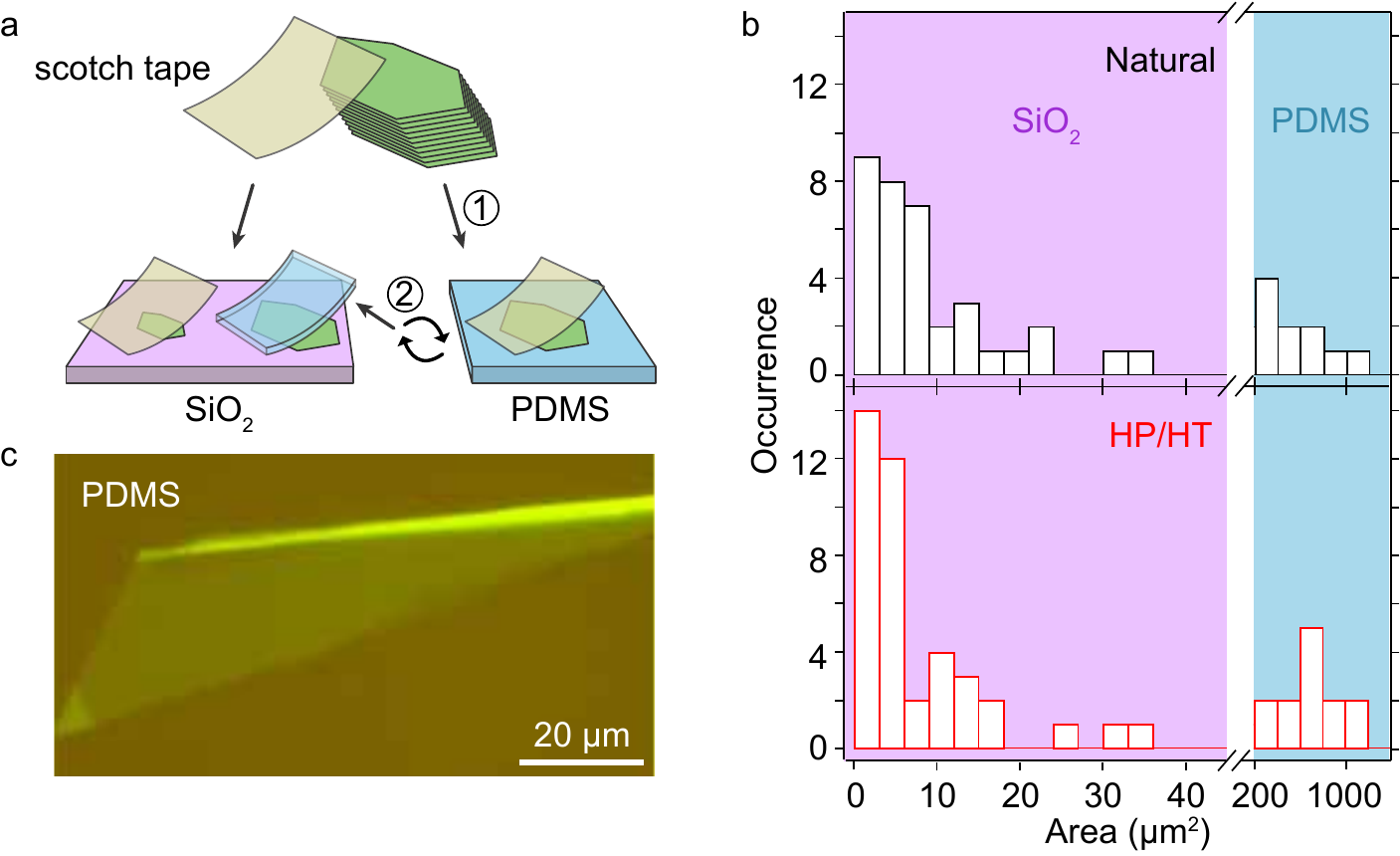}
 \caption{\label{fig1}(a) Schematics of the process for exfoliation with scotch-tape towards a SiO$_2$ surface and PDMS. (b) Occurrence of MoS$_2$ flakes with less than four layers in thickness, obtained by scotch-tape exfoliation on SiO$_2$ (light pink-shaded regions) and stamping on PDMS (blue-shaded regions), in the case of a natural source of MoS$_2$ (top) and the HP/HT-MoS$_2$ (bottom) for both scotch-tape and PDMS. 45 and 52 flakes have been measured for the two sources of MoS$_2$. (c) Optical micrograph of a large bi-layer MoS$_2$ flake obtained by exfoliation on PDMS.}
 \end{center}
\end{figure*}

\begin{figure*}[!ht]
  \begin{center}
  \includegraphics[width=122.2mm]{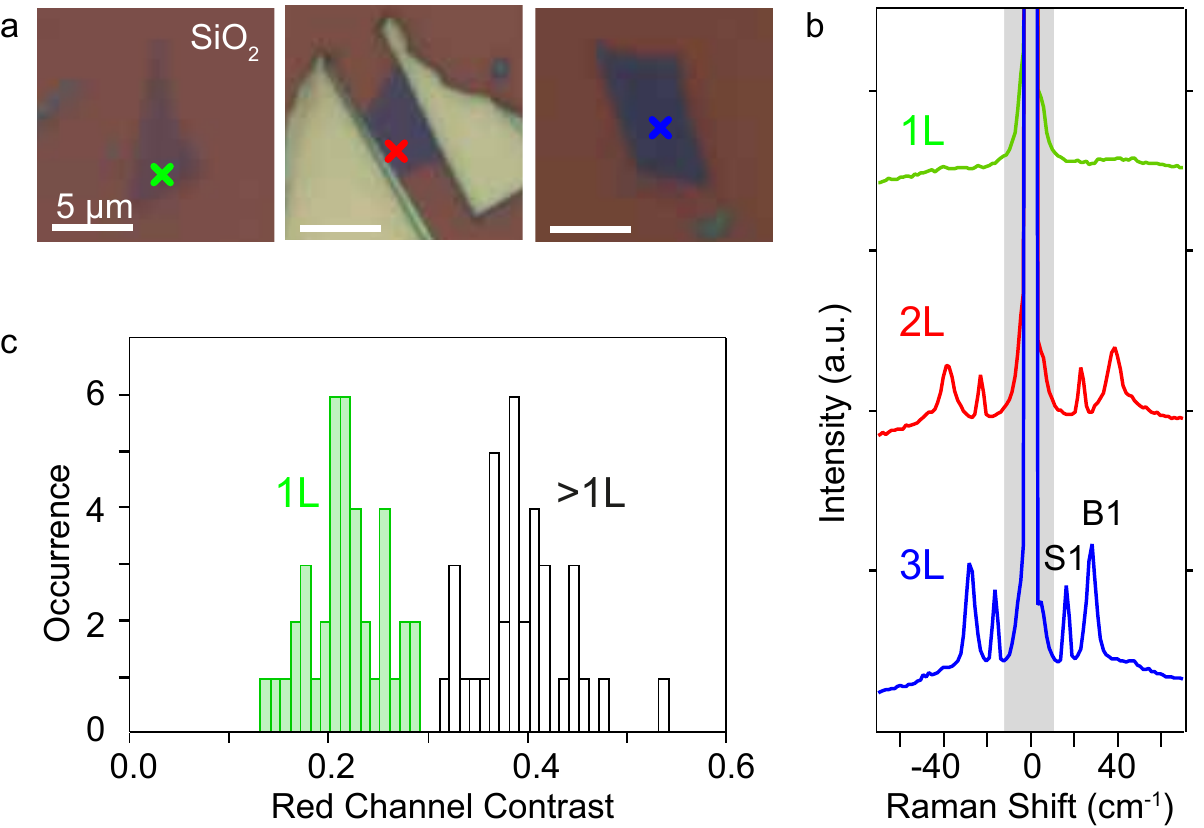}
  \caption{\label{fig2}(a) Optical micrographs of (from left to right) single- (1L), bi- (2L), and tri- (3L) layer MoS$_2$ flakes exfoliated on SiO$_2$ with scotch-tape. (b) Raman spectra of the inter-layer shear (S1) and breathing (B1) modes measured at the locations marked with a cross in (a). The three spectra are vertically-shifted for clarity; the grey-shaded area corresponds to the stop band of the notch filter (within which the measured intensity is not informative). (c) Occurrence of all MoS$_2$ 1L flakes (green bars) showing a characteristic optical contrast relative to SiO$_2$ in the red channel in numerical images (see Methods and Materials). For comparison, some flakes with more than one layer in thickness (black bars), showing a higher optical contrast, are shown. 75 flakes have been measured in total. Green bars signal single-layer flakes, as ascertained with Raman spectroscopy (see Methods and Materials), and black bars signal flakes with more than one layer in thickness.}
  \end{center}
\end{figure*}

Here, we report on point defects that induce defect-bound excitons in single-layers of MoS$_2$. The MoS$_2$ is obtained from two different sources of bulk crystals. We use a natural crystal and a synthetic crystal, prepared at high pressure and high temperature (denoted HP/HT in the following). We find that the latter kind of bulk MoS$_2$ hosts specific defects, and holds promise for improved control of the structure of MoS$_2$ in the future. We discriminate the electronic doping, mechanical strain, and defect-induced exciton localisation by combining Raman spectroscopy, photoluminescence mapping, scanning tunneling microscopy (STM), and density functional theory (DFT) calculations. We also discriminate the influence of extrinsic and intrinsic effects by addressing samples transfered on silica and on hexagonal boron nitride (\textit{h}-BN).\\

\textbf{Preparation of MoS$_2$ few- and single-layers.} The natural MoS$_2$ bulk crystals (typical size, 2~mm) used in this study are provided by SPI-supplies. The second kind of bulk MoS$_2$ we used are prepared at NIMS, Tsukuba, following a slow cooling process from a molten state attained under high pressure. Here, MoS$_2$ (99.9\% pure), supplied by Kojund Chemical Laboratory Co. Ltd., is encapsulated in a \textit{h}-BN capsule and brought to 5~GPa and 1800$^\circ$C for 20~min by using a belt-type high pressure apparatus. The sample is then cooled down to room temperature at a rate of 0.8$^\circ$C/min. After releasing the pressure, the MoS$_2$ crystal is recovered by crushing the \textit{h}-BN capsule. The crystal size after this process is typically 1~mm.

Mechanical exfoliation of MoS$_2$ was achieved with two processes (Figure~\ref{fig1}a). In the first one, a macroscopic MoS$_2$ grain attached onto scotch-tape is thinned down with repeated scotch-tape exfoliation. Next the surface of the MoS$_2$-covered tape is stamped onto a SiO$_2$ wafer.\cite{Novoselov} Irrespective of the source of MoS$_2$, the typical area is of the order of few 1 to 10~$\mu$m$^2$ (Figure~\ref{fig1}b). The other process, using a polydimethylsiloxane (PDMS) host support\cite{Castellanos} instead of SiO$_2$, substantially increases the area of the exfoliated flakes, in the few 100 to 1,000~$\mu$m$^2$ range (Figure~\ref{fig1}b). Figure~\ref{fig1}c displays a photograph of one of the largest flakes (a bi-layer one) that we exfoliated, among the several tens we have prepared. We note that the transfer processes that we used are dry-process that are not expected to alter the atomic structure of the individual MoS$_2$ layers. The use of PDMS limits the amount of contaminants left on MoS$_2$,\cite{Castellanos} as observed by atomic force microscopy and electron energy loss spectroscopy performed in a scanning transmission electron microscope (see Supporting Information, Figures~S1,S2).

\begin{figure*}[!ht]
  \begin{center}
  \includegraphics[width=165.0mm]{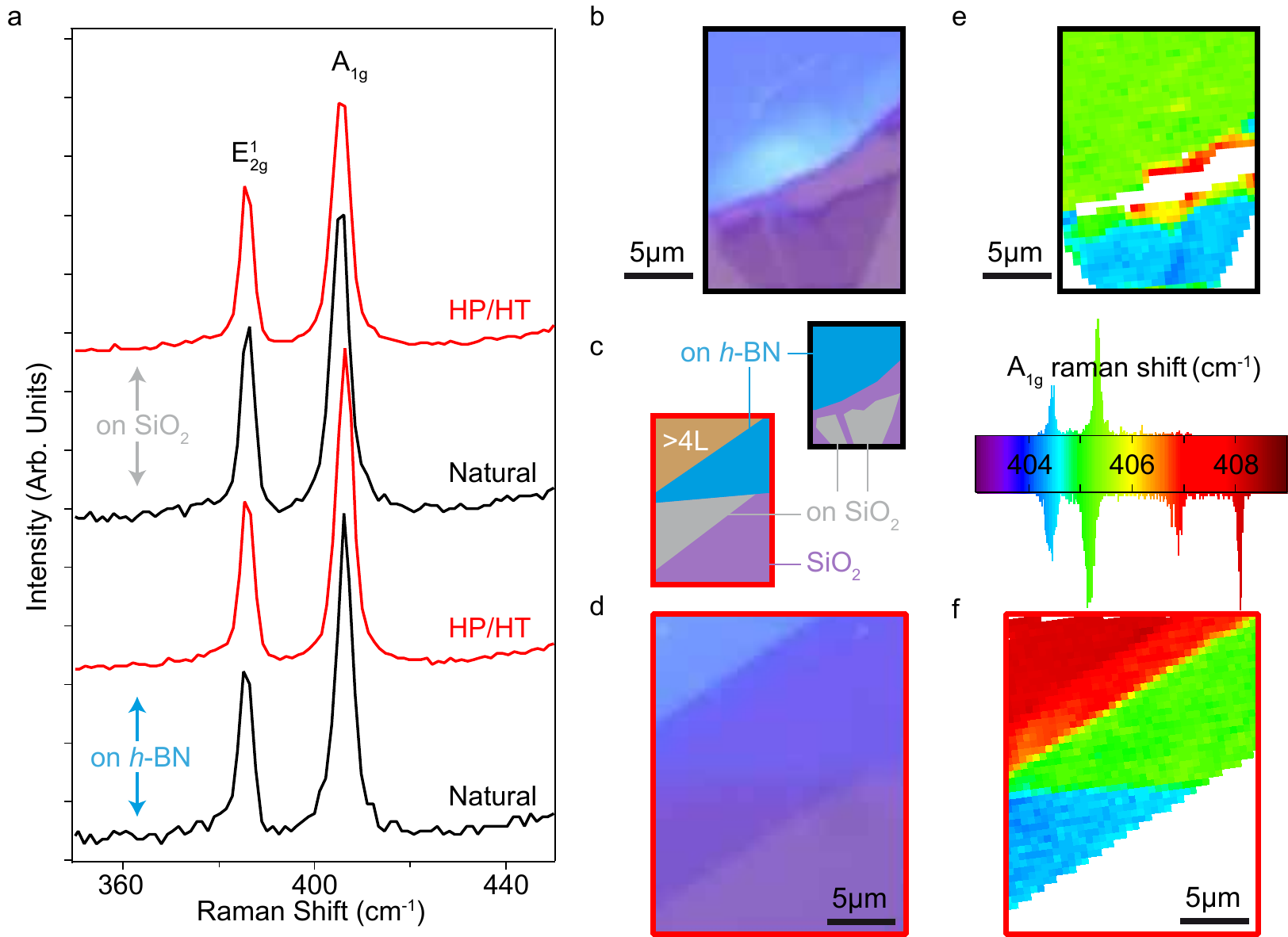}
  \caption{\label{fig3}(a) Raman spectra (532~nm-wavelength laser) for MoS$_2$ single-layers exfoliated from a natural crystal (black) and from a HP/HT source (red), on SiO$_2$ and \textit{h}-BN. (b,c,d) Optical micrographs for MoS$_2$ exfoliated from the two kinds of crystals. The cartoons (c) clarify the stacking of MoS$_2$ on \textit{h}-BN and SiO$_2$. (e,f) Raman maps of the position of the A$_{1g}$ mode for the region corresponding to (b,d), for the two kinds of MoS$_2$ samples. The distribution of the mode position is shown for the top and bottom maps. The thick black and red frames in (b-f) refer to two MoS$_2$ sources, natural and HP/HT.}
  \end{center}
\end{figure*}

As a preliminary step, we present the way we determine the thickness of the flakes. Our approach is to establish the correspondence between optical contrast, which varies with the number of layers and optical wavelength in a rather complex manner\cite{Castellanos_b,Benameur} (Figure~S3), and an unambiguous independent determination of the number of layers with Raman spectroscopy. We track the occurrence and position of the shear (S1) and breathing (B1) interlayer vibrational modes.\cite{Plechinger,Zhao} Figure~\ref{fig2}a shows three MoS$_2$ flakes exfoliated with scotch-tape onto SiO$_2$. Their Raman spectra are different, with the single-layer MoS$_2$ readily identified by the absence of the B1 and S1 modes, while these two modes are found in bi- and tri-layers, and are stiffer in the latter case (Figure~\ref{fig2}b). We find that in the red-channel of the digital optical images, the contrast is the lowest, of 0.22$\pm$0.08 for a single-layer (Figure~\ref{fig2}c). This characteristic contrast value is used as a criterion for fast identification of single-layers. The remaining of the paper is focused on single-layers.

Using PDMS stamps where single-layers are first identified, we then transferred MoS$_2$ onto two kinds of substrates. The first one is SiO$_2$, and the second one is \textit{h}-BN, which has been exfoliated onto SiO$_2$ beforehand. In both processes, the interface between MoS$_2$ and the substrate has not been exposed to the polymer (PDMS), hence pristine MoS$_2$/support interfaces are formed.

\textbf{Strain and electronic doping from vibrational spectroscopy.} Figure~\ref{fig3}a shows characteristic Raman spectra zoomed in the low wavenumber region, featuring intralayer shear (E$^1_{2g}$) and breathing (A$_{1g}$) modes (which are stiffer than the interlayer modes addressed above, for which the bond strength is much lower), for the two sources of MoS$_2$ on the two substrates. To better highlight the differences between the four possible stacks (on the two substrates, for each of the two sources), we mapped the position of the A$_{1g}$ mode (determined by Lorentzian fits of the corresponding peak), which is especially sensitive to electron-phonon coupling effects,\cite{Chakraborty} across an area corresponding to the optical micrographs shown in Figures~\ref{fig3}b-d. The result is shown in Figures~\ref{fig3}e,f. The most obvious difference is the correlation between the position of the A$_{1g}$ mode and the nature of the substrate: a blue-shift, of 1.0$\pm$0.1 and 0.8$\pm$0.1~cm$^{-1}$ from the SiO$_2$ to the \textit{h}-BN substrate, is observed for the natural and HP/HT sources respectively.

\begin{figure*}[!ht]
  \begin{center}
  \includegraphics[width=110.6mm]{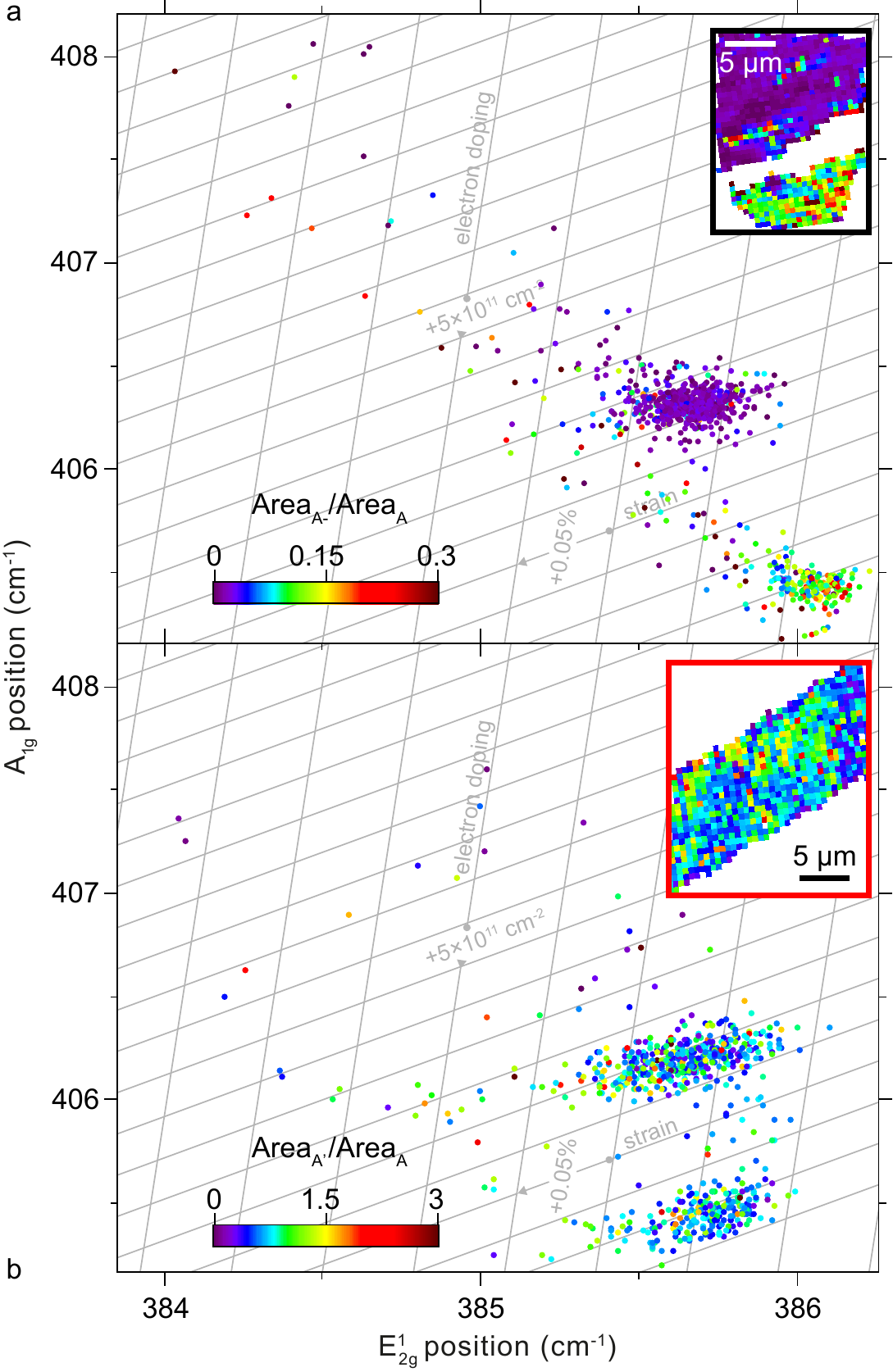}
  \caption{\label{fig3bis}Positions of the A$_{1g}$ and E$^1_{2g}$ modes of the Raman spectra (532~nm-wavelength laser), each point corresponding to a point in the maps shown in Figures~\ref{fig3}e,f. The grid of the strain \textit{vs} electronic doping has increments of 0.05\% and 5$\times 10^{11}$~cm$^{-2}$. (a) and (b) correspond to the two MoS$_2$ sources, natural and HP/HT respectively. Each point is coded with a colour corresponding to the ratio of areas of the two contributions to the main excitonic feature in the photoluminescence spectra, shown in the insets (see Figure~\ref{fig4}). Inset: Spatial dispersion of area ratio.}
  \end{center}
\end{figure*}

These blue shifts may be caused by mechanical strain\cite{Conley,Castellanos_c,Parkin} and/or by electron doping,\cite{Chakraborty} translating the anharmonicity of the interatomic potentials and the effect of the electron-phonon interaction respectively. The energy of the A$_{1g}$ and E$^1_{2g}$ modes have characteristic variations with each of the effect. A strain \textit{vs} doping graph can hence be extracted from the maps of the A$_{1g}$ (Figures~\ref{fig3}e,f) and E$^1_{2g}$ Raman shifts --- a two-dimensional space is constructed with the positions of the A$_{1g}$ and E$^1_{2g}$ modes as principal axis.\cite{Michail} Figure~\ref{fig3bis} shows such a graph for the two samples. Disregarding at this stage the colours of the points (which will be discussed later in the light of the photoluminescence measurements), for both samples we find two groups of points, each corresponding to MoS$_2$ on \textit{h}-BN (greater A$_{1g}$ positions) and on SiO$_2$. The trend is similar for both sources of MoS$_2$, suggesting that the observations mostly point to an extrinsic effect, namely the nature of the substrate. The electron doping level is larger by 2.5 and 2.0$\times 10^{12}$~electrons/cm$^{-2}$ on SiO$_2$, for the natural and HP/HT sources respectively. Substrate-induced doping is a known phenomena, which was ascribed, in other two-dimensional materials,\cite{Lu_b,Bao,Kretinin} to charged impurities in SiO$_2$ that are absent in \textit{h}-BN. These charged impurities effectively dope MoS$_2$ with electrons, and the observed doping level is consistent with a previous observation\cite{Buscema} --- they hence represent extrinsic defects.

Figures~\ref{fig3bis}a,b also reveal that the transfer process can generate non-uniform strains to small extent. The two clusters of points from both images are scattered within typically 0.05 to 0.1\%. Besides, in the case of Figure~\ref{fig3bis}a (natural source of MoS$_2$), a strain difference of 0.05 to 0.1\% is found between the two clusters of points, corresponding to the top and bottom part of the optical image (on \textit{h}-BN and SiO$_2$ respectively). We do not find such a difference in Figure~\ref{fig3bis}b (HP/HT-MoS$_2$). The observed differences are not systematic, and we believe that they point to slightly different mechanical efforts exerted during the preparation and/or different \textit{h}-BN thicknesses for the two samples, rather than from, \textit{e.g.}, internal strain induced by defects.\cite{Parkin}\\

\begin{figure*}[!ht]
  \begin{center}
  \includegraphics[width=144.9mm]{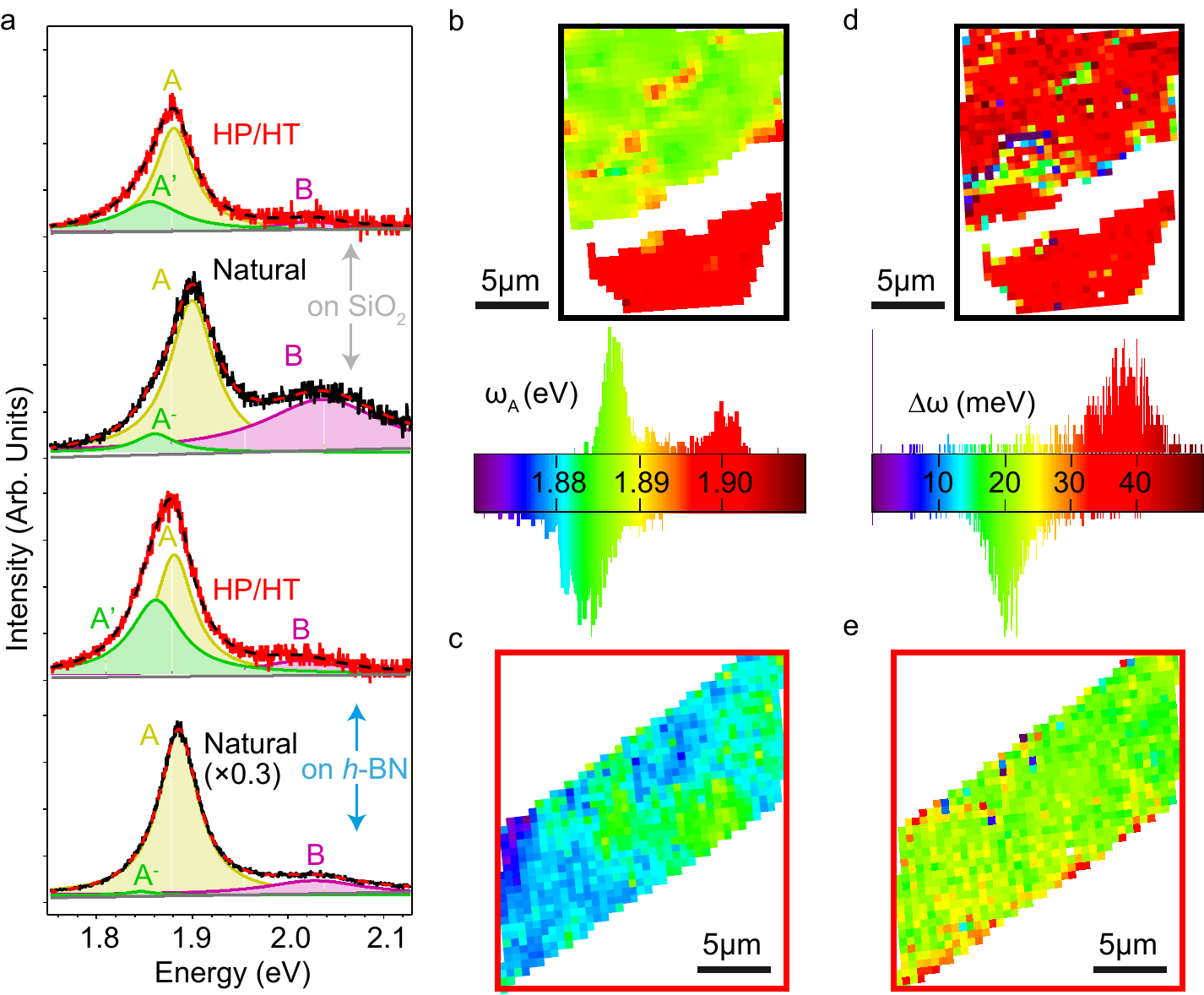}
  \caption{\label{fig4}(a) Room temperature photoluminescence spectra for single-layer MoS$_2$ prepared by exfoliation from natural (black) and HP/HT (red) crystals, deposited on the surface of SiO$_2$ (top) and \textit{h}-BN (bottom). The spectra have been corrected for interference effects associated to the presence of the MoS$_2$/\textit{h}-BN, MoS$_2$/SiO$_2$, \textit{h}-BN/SiO$_2$, and SiO$_2$/Si interfaces. The spectra are fitted with three Lorentzian components, respectively corresponding to the lowest energy direct transition, for the exciton (A), the trion for natural MoS$_2$ (A$^-$) or the defect-bound exciton for HP/HT MoS$_2$ (A'), and the second lowest energy direct transition for the exciton (B). The dotted lines are the best fits to the data. (b-e) Maps of energy of the A component, $\omega_\mathrm{A}$ (b,c), and of the difference in energy $\Delta\omega$ of the A$^-$ and A (d) and A' and A (e) components, for the same area as in Figures~\ref{fig3}b,d, for natural (thick black frames, b,d) and HP/HT (thick red frames, c,e) MoS$_2$ single layers.}
  \end{center}
\end{figure*}

\textbf{Excitonic complexes in presence of strain, electronic doping, and defects.} Both electronic doping level and strain influence the excitonic properties of MoS$_2$.\cite{Mak_b,Mouri,Nan,Pei,Conley} To address these effects we performed photoluminescence measurements at room temperature with a 532~nm laser excitation and low power (see Materials and Methods) for both sources of MoS$_2$ and both substrates. Figure~\ref{fig4}a displays characteristic spectra corrected from optical interference effects (special care needs to be paid to these corrections, see Supporting Information, Figures~S3,S4). As expected with an excitation wavelength of 532~nm, two main excitonic peaks are observed, each corresponding to a different transition involving one or the other spin-polarized valence band.\cite{Steinhoff} In the following, we will focus on the lowest energy peak, and to start with, we address the natural source of MoS$_2$. This peak actually comprises two components. They are separated by typically 40~meV and have a full-width at half maximum of several 10~meV dominated by electron-phonon coupling effects.\cite{Dey} They correspond to a neutral (A) exciton and a charged (A$^-$) exciton --- a trion.\cite{Mak_b,Sercombe} The latter is more prominent when the electronic doping is higher.\cite{Mak_b}

As discussed in Ref.~\citenum{Mak_b}, the ratios of areas of the two peaks as inferred from photoluminescence (Figure~\ref{fig3bis}a) characterise the level of electron doping; we estimate it to be typically of the order of several 10$^{12}$~electrons/cm$^{-2}$. The inset of Figure~\ref{fig3bis}a reveals a distinctive trion \textit{vs} exciton population whether MoS$_2$ lies on SiO$_2$ or \textit{h}-BN. The ratios of A$^-$ to A areas, typically 0.01-0.02 and 0.1 respectively on these two substrates, are consistent with changes of electronic doping levels, due to charged impurities in SiO$_2$, found in the analysis of the Raman data (of the order of a few $10^{12}$~cm$^{-2}$).

The position of the two peaks (Figures~\ref{fig4}b,S5) is changing by 12~meV whether MoS$_2$ lies on SiO$_2$ or \textit{h}-BN. As Raman spectroscopy suggests, this is a result of the preparation process causing a spatial strain variation, a compression on SiO$_2$ relative to the case on \textit{h}-BN, by about 0.1\%. The magnitude of the strain-induced energy shift fits with that corresponding to previously reported strain-induced electronic band-gap change.\cite{Conley}

Let us now turn to the photoluminescence signatures in case of HP/HT-MoS$_2$. In this case also, we find that the main excitonic feature does not consist of a single component. While the above-discussed energy difference between the two componenents was about 40~meV for natural MoS$_2$, consistent with the expected trion binding energy corresponding to electron doping levels of the order of few $10^{12}$~electrons/cm$^{-2}$,\cite{Mak_b}  here the two components are separated by a substantially lower energy difference (20~meV), regardless of the substrate (Figure~\ref{fig4}c). Such an energy difference cannot correspond to a trion under the influence of strain or electronic doping: the variations of strain and electronic doping in our samples are in the range of few 0.1\% and 10$^{12}$~cm$^{-2}$ respectively, which have only marginal influence on the binding energy of the trion (few 1~meV or below).\cite{Wang_c,Mak_b} What is then the nature of this low-energy emission?\cite{note_on_components} Its spectral weight is globally high and strikingly, unlike the A$^-$ feature for natural MoS$_2$, does not corelate with the kind of substrate, and corresponding doping level revealed by Raman spectroscopy (Figure~\ref{fig3bis}b). This is at variance to the behaviour expected for trions.

A rational explanation for this low-energy feature (in the case of HP/HT MoS$_2$) is that it relates to a defect-bound exciton. Defect-bound excitons were previously invoked in MoS$_2$ and attributed to sulfur vacancies, di-vacancies, and metal vacancies.\cite{Tongay,Chow} While they were found to be associated with a binding energy of the order of 100~meV, here we find a binding energy of 20~meV. The limited variations of strain or electronic doping in our samples do not allow us to reveal a possibly different influence of these effects on the A', A$^-$ and A features. As we will see, samples from the HP/HT source comprise a larger amount of defects. In the following, we devise on the nature and density of these defects using additional probes.\\

\textbf{The nature of the point defects.} A large variety of defects has been considered in MoS$_2$, including sulfur vacancies,\cite{Kim,Tongay,Qiu,Komsa} substitutional atoms replacing either the metal or the sulfur atom,\cite{Dolui,Kim,Tongay,Qiu,Noh,Komsa} and individual atoms (the electrodonor alkali atoms) adsorbed onto the surface.\cite{Dolui} Only the latter kind has been reported to be associated to shallow donor levels, that could account for usually reported $n$-doping in single-layer MoS$_2$ at room temperature. The chemical analysis of the starting material in the HP/HT process does not seem compatible with the presence of alkali atoms, though. On the contrary, based on this analysis, potential candidates as impurities are iron and carbon prominently, or boron and nitrogen from the capsule used to seal the MoS$_2$ during the HP/HT treatment.

\begin{figure*}[!ht]
  \begin{center}
  \includegraphics[width=117.5mm]{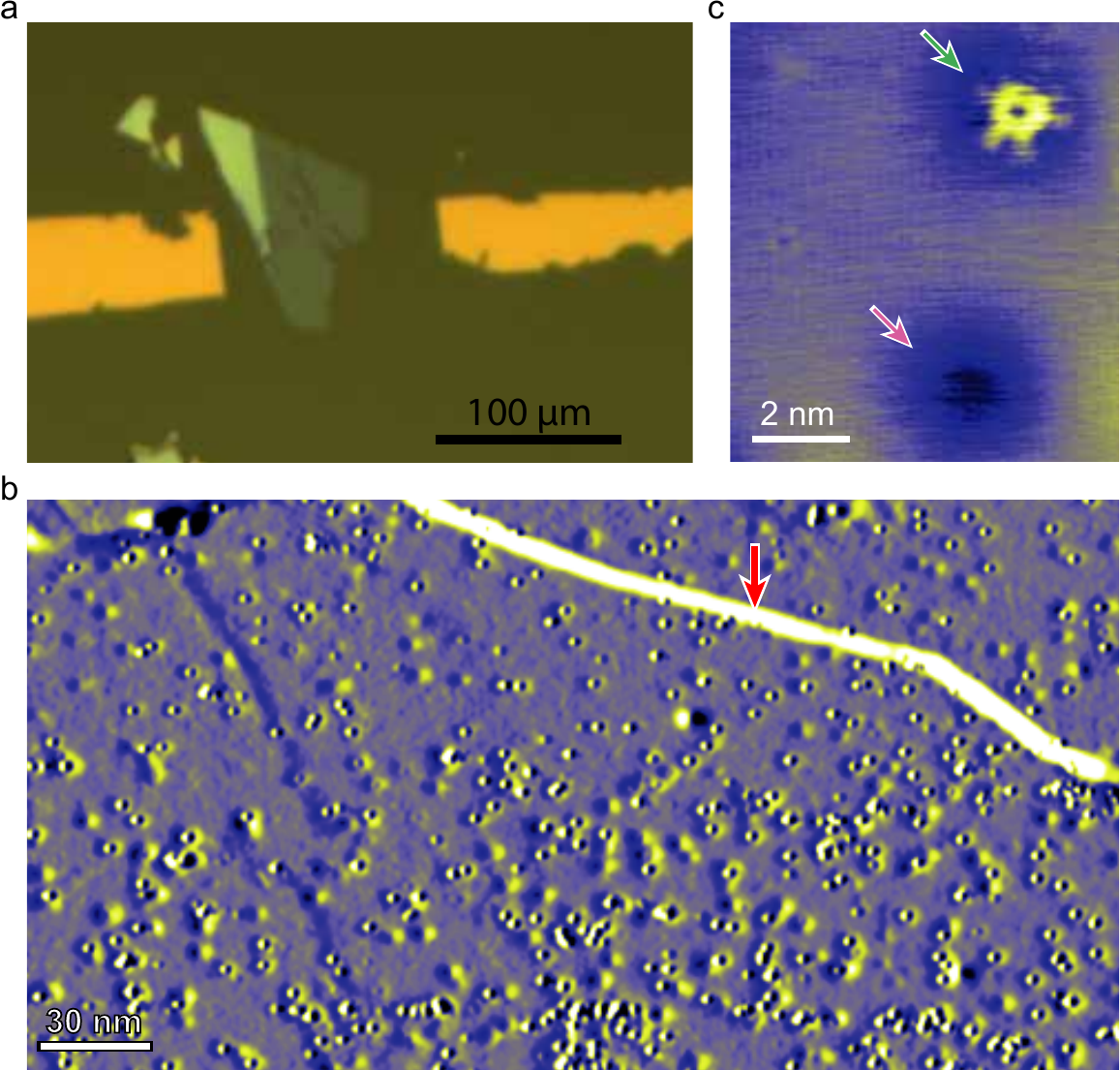}
  \caption{\label{figSTM}(a) Optical micrograph of the five-layer HP/HT MoS$_2$ deposited on graphene/SiC, with gold markers. (b) STM image measured with a bias voltage $V_\mathrm{b}$=-2~V and a tunneling current of $I_\mathrm{t}$=0.2~nA. The image shows the derivative of the apparent height as function of the horizontal spatial coordinate to enhance the contrast of the atomic-size defects. The red arrows points to an atomic step edge of the substrate, with a height of 0.75~nm. (c) STM topograph ($V_\mathrm{b}$=-2~V $I_\mathrm{t}$=0.2~nA) close-up view on some defects. The green and pink arrows point to two kinds of defects.}
  \end{center}
\end{figure*}

The impurity levels in MoS$_2$ are too low to be reliably assessed with standard macroscopic chemically-sensitive probes such as X-ray photoelectron spectroscopy.\cite{Addou} High resolution microscopy circumvents this issue, by addressing the defects individually. We used STM for this purpose, as implemented in a ultra-high vacuum environments that limits spurious interactions of the defects with \textit{e.g.} small molecules. Very few reports in the literature have in fact been devoted to STM measurements on single- or few-layer flakes. Mostly, this is due to the small size (few 1 to 10 $\mu$m$^2$) of the exfoliated flakes, which if deposited on a non-conductive substrate, must be electrically contacted with finely designed electrodes. The observation of such small features with a short-field-of-view-technique as STM is obviously very laborious. This is probably why most reports since 20 years rely on cleaved bulk MoS$_2$.\cite{Abe,Murata,Park,Addou,Addou_b,McDonnell,Lu} Two workarounds have been recently implemented: one, taking benefit of large-area growth of MoS$_2$ on graphite,\cite{Huang,Zhou} and the other the strong adhesion of MoS$_2$ exfoliated on a gold surface.\cite{Magda} As far as we are concerned, we chose an alternative strategy and once more exploited PDMS exfoliation (Figure~\ref{fig1}a) which yields large flakes of sizes approaching 100~$\mu$m, using as a host substrate a (conductive) graphene-covered silicon carbide surface. To ease the localisation of the (few-layer) flake we further deposited micrometer-sized gold markers (Figure~\ref{figSTM}a, see Methods and Materials).

This rather advanced sample preparation allows to image single defects with STM (yet it should be noted that the measurements are in no way straightforward). A high density of defect is observed (Figure~\ref{figSTM}b), of the order of 1$\times$10$^{12}$~cm$^{-2}$, varying from 0.6 to 4$\times$10$^{12}$~cm$^{-2}$ from one place to the other. A spatially inhomogeneous distribution of defects was already quoted in previous STM analysis from MoS$_2$ samples.\cite{Addou_b,Addou,Vancso} The density we find on the HP/HT sample is larger than the one observed on samples prepared by exfoliation of natural molybdenite, which is in the few 10$^{11}$~ cm$^{-2}$ range\cite{Addou_b,Addou} or less\cite{Lu} (3.5$\times$10$^{10}$~cm$^{-2}$). Conversely, a much larger density of defects (from 5$\times$10$^{12}$~cm$^{-2}$ to 5$\times$10$^{13}$~cm$^{-2}$) has been reported for MoS$_2$ prepared by exfoliating synthetic crystals.\cite{Vancso}

In the HP/HT MoS$_2$ we find two prominent populations of point defects, which appear as a bright feature and a depression respectively (Figure~\ref{figSTM}c). Depression-like defects of the same extension (1-2~nm) or slightly larger (in the few nanometer-scale) have been reported previously.\cite{Inoue,Abe,Addou,Addou_b,Lu}. Among them, one appears as a depression at negative tip-sample bias as in our observations, and is a characteristic defect in natural MoS$_2$ that is ascribed to missing S-Mo-S fragments located either in the top or in a buried MoS$_2$ layer.\cite{Addou_b} The second kind of defect (the bright one) has not been observed in natural MoS$_2$ samples, and is hence generated during the preparation of the HP/HT sample. It has a characteristic shape resolved with sharp STM tips, consisting in a ring with three pairs of radial legs. The size of the ring is typically 0.7~nm. Defects featuring a ring shape in STM have also been reported previously\cite{Abe,Murata} and were ascribed to alkali atoms adsorbed on the surface. Nor the HP/HT process neither the ultra-high vacuum chamber where the STM measurements were performed seem to yield such adsorbates, though.

\begin{figure}[!ht]
  \begin{center}
  \includegraphics[width=82.2mm]{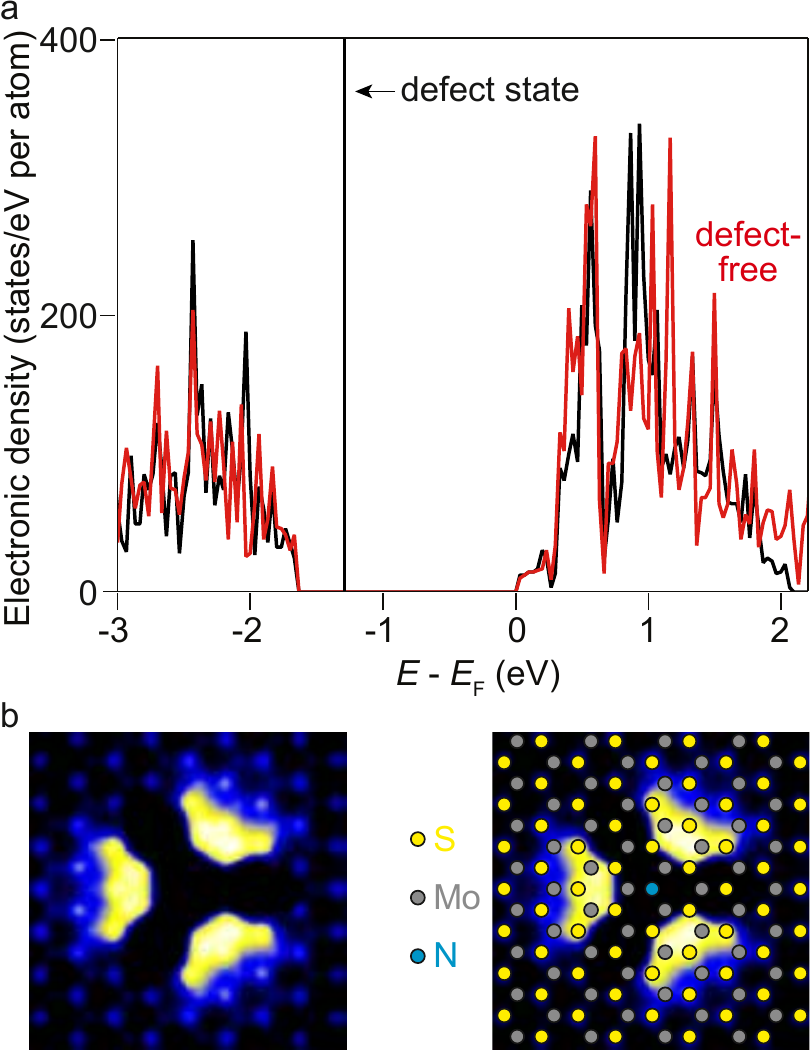}
  \caption{\label{figDFT}(a) Electronic density of states as a function of electron energy with respect to the Fermi level, for a nitrogen atom substituting a sulfur atom in single-layer MoS$_2$, and for defect-free MoS$_2$ (red). The spectra have been shifted horizontally to match the bottoms of the conduction band. The arrow points the position of a very sharp defect state. (b) Corresponding simulated STM image with Fermi level located at the bottom of the conduction bond and -2~V tip-sample bias, with the STM tip 4~\AA\, from the surface, with (right) and without (left) the atomic structure overlaid.}
  \end{center}
\end{figure}

Both electronic and structural information contribute to STM images. To devise on the nature of the defects, DFT simulations provide key insights to interpret the observed STM contrasts. The comparison of experimental STM images with spatially-resolved information provided by DFT is a well-established approach to study defects. To our knowledge such comparison has not been made in the case of defects in MoS$_2$ beyond the case of lattice vacancies.\cite{Inoue,Vancso} We computed the stable configurations of five defects corresponding to the impurities that are detected in the chemical analysis of the raw MoS$_2$ or present in the \textit{h}-BN capsule used in the HP/HT process. This includes a sulfur vacancy, a molybdenum atom substituted by an iron atom, an a sulfur atom substituted by a carbon, a nitrogen, and a boron atom. Each of these defects are associated with electronic states inside the bandgap of MoS$_2$ (here expectedly close to the bulk value of 1.3~eV) or close to the bandgap edges (see Figures~\ref{figDFT}a,S7). The sample bias of -2.0~V corresponds to electrons tunneling from the sample to the tip, in an energy window of 2.0~eV below the MoS$_2$ Fermi level, which is presumably located close to the conduction band minimum. It is thus expected that the defect electronic states within the bandgap have significant contribution compared to the valence band, given that they correspond to a lower tunnel barrier. We simulated STM images by taking into account the STM tip (see Materials and Methods) in presence of the different defects. The results are shown in Figure~S8, and for one specific defect (nitrogen atom substituting a sulfur atom) in Figure~\ref{figDFT}b. For the latter defect, the simulated STM image is in rather good agreement with the experimental one, despite the significant difference in spatial resolution, which is higher in the simulations. Indeed in the experiment, an {\AA}ngstr\"{o}m-scale instability of the scanning tip is observed (as shown by the occurence of horizontal stripes at the defect location in Figure~\ref{figSTM}c), and the tip's shape presumably deviates from the ideal pyramidal shape assumed in the calculations. We consider this as the reason why the three brilliant lobes observed in the simulated image appear as a circle in the experimental image. Beyond this, the main features compare very well for the N substitutional defect: the size of the lower-intensity central feature match, and the three pairs of legs appearing in the experimental image seem reminiscent of the three lobes found in the simulations. Based on this comparison we propose that the ring-shaped defects we observed correspond to nitrogen atoms having replaced sulfur ones during the HP/HT sample preparation (and originating \textit{e.g.} from the \textit{h}-BN capsule used in this process).\\

\begin{figure*}[!ht]
  \begin{center}
  \includegraphics[width=105.6mm]{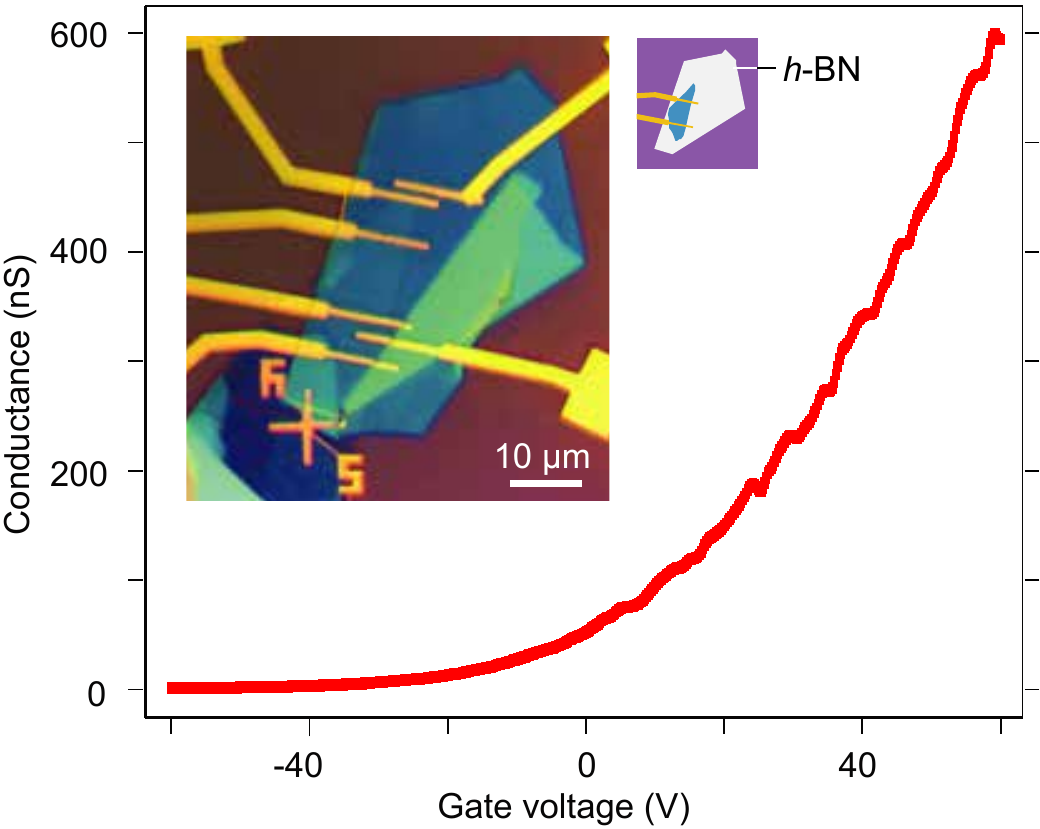}
  \caption{\label{fig5}Room temperature conductance measured under vacuum in a single-layer transistor based on HP/HT MoS$_2$, as a function of gate voltage, with a source-drain bias of 0.2~V. The inset shows an optical micrograph of a single-layer device, which is transferred on \textit{h}-BN (exfoliated on SiO$_2$) and contacted with Au electrodes. The two contacts used for measuring the conductance are shown in the cartoon.}
  \end{center}
\end{figure*}

\textbf{Field-effect transistor based on HP/HT MoS$_2$.} Single-layer MoS$_2$ prepared under HP/HT conditions was finally integrated into a field effect transistor with electrostatic gating from the back-side, in which direct contact with SiO$_2$ was avoided by a \textit{h}-BN buffer layer (Figure~\ref{fig5}). Accordingly a low amount of charged impurities is expected in the vicinity of MoS$_2$. Consistent with previous reports, we find that the conduction properties are improved under vacuum (compared to ambient pressure), presumably due to the desorption of species acting as charged impurities.\cite{Lembke} We only observe the blocked state of the transistor and the regime of electron conduction (and not the hole conduction regime) in the source-drain current \textit{vs} gate-voltage characteristic (Figure~\ref{fig5}).

The transport properties overall show very typical semiconducting properties which match those found in similar devices based on natural MoS$_2$. We estimate the threshold voltage to be at a gate voltage of 10~V. The mobility from the gating curve is estimated to be 2~cm$^2$V$^{-1}$s$^{-1}$ (the device in the on-state has not reached saturation in the range of applied gate-voltage, so this value is a lower-estimate). These two values are similar to those found in devices using the same geometry based on natural MoS$_2$, and more recently in similar field-effect transistor architecture featuring single-layer MoS$_2$ synthesised by chemical vapor deposition transfered onto \textit{h}-BN.\cite{Joo,Joo_b} In all these works and ours, we stress that the Schottky barriers at the Au/MoS$_2$ junctions under the source and drain electrodes play a dominant role in the low-value-mobility obtained from the two-probe measurement; in other words the defects that are present in the MoS$_2$ channel are not limiting transport in this configuration.

\section*{Conclusions}

Using Raman spectroscopy, photoluminescence spectroscopy, scanning tunneling microscopy, density functional theory, and electronic transport measurements, we addressed the optoelectronic properties of MoS$_2$ single-layers prepared by exfoliation from two different sources of bulk material --- a natural one, and another one prepared under high-pressure and high-temperature (HP/HT) conditions. The latter preparation process opens the route to the control of the structure of MoS$_2$, in terms of intentional generation of otherwise inaccessible defects and possibly in the future as well in terms of superior quality, namely increased single-crystal size and lower defect (\textit{e.g.} vacancies) concentration as was achieved with \textit{h}-BN.\cite{Watanabe} This holds promise for close-to-ideal support for other two-dimensional materials\cite{Lu_c} and high-performance optoelectronic devices. Natural and HP/HT both have a substantial electron-type doping, of the order of 10$^{12}$~cm$^{-2}$, which is stronger on SiO$_2$ substrates than on \textit{h}-BN, due to a lower amount of extrinsic charged impurity in the latter case. Additional defects are present in HP/HT MoS$_2$. We argue that they lead to defect-bound excitons, with a few 10~meV binding energy. We propose that these defects are nitrogen atoms substituting sulfur atoms. Exploring the nature of the localisation potential associated to the defects, and its effects on the coupling to the electromagnetic field, will provide valuable insights to understand light-matter coupling in transition metal dichalcogenides. In addition, the defect-bound exciton we discover may couple coherently with the neutral exciton. Stronger coupling and larger coherent times than reported in MoSe$_2$ at low temperature between excitons and trions\cite{Singh} might result from the weak trapping of the defect-bound exciton. On a general note, our work also sheds light on the influence of defects on the optoelectronic properties of these two-dimensional materials and their interplay with internal and external (force, electric, optical) fields.

\section*{Materials and methods}

\textbf{Mechanical exfoliation.} Si/SiO$_2$ substrates with 285~nm-thick oxide were cleaved into 1~cm $\times$ 1~cm pieces. They were cleaned using acetone and isopropyl-alcohol, followed by a dry nitrogen blow. The substrates were subjected to a oxygen plasma for 3 to 4 minutes. For mechanical exfoliation, small pieces of MoS$_2$ crystal were placed on the scotch tape followed by repeated exfoliation. The tape was then stamped on to the Si/SiO$_2$ substrate, then it was then heated for 15-30~s at 80-100$^\circ$C on a hot plate. The tape was then gently removed from the substrate at an angle of 60 to 70$^\circ$. A similar process was followed on the PDMS substrate, without the application of heat. For the MoS$_2$/\textit{h}BN heterostructures, a PDMS stamping method was used.\cite{Castellanos}

\textbf{Raman spectroscopy and photoluminescence measurements.} Raman spectroscopy and photoluminescence were acquired with 532~nm Nd:YAG laser using a commercial confocal WITEC spectrometer at room temperature under ambient condition. The laser spot size was $\sim$1~$\mu$m. The signal was collected through a 50$\times$ objective with a numerical aperture of 0.75. For the Raman spectra, the power was kept at 300~$\mu$W to avoid damage due to laser-induced heating in MoS$_2$ flakes. The signal was integrated for 2~s after being dispersed by a 1800~lines/mm grating. For photoluminescence measurements, a low power of 8~$\mu$W (see Supporting Information and Figure~S6) was used with a grating of 600~lines/mm. The photoluminescence spectra were taken with an integration time of 30~s to improve signal-to-noise ratio, and the spatially-resolved photoluminescence maps were taken with an integration time of 5~s.

Optical images have been acquired in a Zeiss microscope equipped with a digital camera "axiomcam 105 colour" device, a tungsten halogen light source and with a magnification of 100$\times$. The white balance has been set to the expected one for an halogen lamp at 3200~K.

To allow for fast determination of the flakes number of layers (less than four layers) across square centimeter-scale surfaces, we determined the value of a representative optical quantity for flakes of known (from Raman spectroscopy) thickness. We chose the contrast of the RBG images, with respect to the surrounding SiO$_2$ surface, in red channel (red channel contrast, RCC) as a relevant quantity, defined as the difference between the signal from SiO$_2$ ($I^\mathrm{R}_\mathrm{silica}$) and from the flake ($I^\mathrm{R}_\mathrm{flake}$), normalised by $I^\mathrm{R}_\mathrm{silica}$: $RCC = (I^\mathrm{R}_\mathrm{silica}-I^\mathrm{R}_\mathrm{flake})/I^\mathrm{R}_\mathrm{silica}$

\textbf{Fabrication and measurement of MoS$_2$ FET device.} The substrate for FET was degenerately-doped silicon with 285~nm of SiO$_2$ on which \textit{h}-BN flakes were exfoliated. The single-layer MoS$_2$ flake was deterministically transferred on \textit{h}-BN using PDMS. Two-step electron-beam lithography followed by metal deposition was used next: in the first step 5~nm of Ti and 65~nm of Au were deposited for the outer pads, and in the second step, 100~nm of Au was used to contact MoS$_2$. The FET measurement was performed at room temperature under vacuum in a probe station.

\textbf{Scanning tunneling microscopy.} The 5-layer MoS$_2$ flake (typically 100$\times$80~$\mu$m$^2$) prepared from the HP/HT source was transfered by PDMS stamping onto graphene grown over the SiC substrate. The average graphene coverage on 6H-SiC(0001), as deduced from Auger electron spectroscopy and STM images, was between one and two layers.\cite{Mallet} Gold markers were evaporated on this substrate and served as alignment marks further helping to locate the MoS$_2$ flake in STM experiments.

STM measurements were performed in an ultra high vacuum (UHV) environment at 300~K using a home-made microscope. The samples were gently outgassed in UHV (typically at 300$^\circ$C for 1~h) before being loaded in the STM setup. The tips were made from mechanically-cut PtIr wires. The data were analysed using the \texttt{WsXM} software.\cite{Horcas}

\textbf{Density functional theory calculations.} Density functional theory calculations were carried out using the Vienna ab initio simulation package \texttt{VASP}, with the projector augmented wave (PAW) approach.\cite{Kresse,Kresse_b} The exchange correlation interaction is treated within the general gradient approximation parametrized by Perdew, Burke and Ernzerhof (PBE)\cite{Perdrew}. Relaxation was performed with a 1$\times$1$\times$1 $k$-point sampling. The energy and forces were converged until 10$^{-4}$eV and 0.01~eV/\AA. Supercells of size (6$\times$6) were used to limit the interaction between image defects associated with the use of periodic boundary conditions. To avoid interaction in the direction perpendicular to the plane of MoS$_2$, a 10~\AA-thick slab of vacuum was used.

\textbf{Simulation of scanning tunneling microscopy images.} The DFT localized-orbital molecular-dynamics code as implemented in \texttt{FIREBALL}\cite{Lewis,Jelinek,Sankey} has been used for the structural relaxation of the different defects in MoS$_2$ considered for STM image calculations. The \texttt{FIREBALL} simulation package uses a localised, optimised minimal basis set,\cite{Basanta} and the local density approximation (LDA) for the exchange and correlation energy following the McWEEDA methodology.\cite{Jelinek} We used a hexagonal (10$\times$10) unit cell for each simulation, in order to reduce the interactions between defects in neighboring cells associated with the periodic boundary conditions. The convergence of the system was achieved using a set of 8 $k$-points in the Brillouin zone, until the forces have reached a value lower than 0.05~eV/\AA. Theoretical simulations of the STM current between the metal tip (placed 4~\AA\, away from the surface) and the sample were based on the non-equilibrium Green's functions technique developed by Keldysh.\cite{Keldysh,Mingo} Within this methodology, the electronic current for an applied voltage $V_\textrm{b}$ at standard tunneling distances can be written as:\cite{Gonzalez}

\begin{equation} \label{stm}
I=\frac{4\pi e^{2}}{h} \int_{E_\mathrm{F}}^{E_\mathrm{F}+eV_\mathrm{b}}\textrm{Tr}[T_{TS}\rho_{SS}(\omega)T_{ST}\rho_{TT}(\omega-eV)]d\omega.
\end{equation}

\noindent where $E_\mathrm{F}$ is the Fermi level, here set at the bottom of the conduction band, $\rho_{TT}$ and $\rho_{SS}$ are the density matrices associated with the subsystem tip and sample and $T_{TS/ST}$ the tip-sample interaction (a detailed discussion can be found elsewhere\cite{Sanchez,Gonzalez}). The $\rho_{TT}$ and $\rho_{SS}$ matrices have been obtained using the hamiltonian obtained after the atomic relaxation. This methodology has already proved to give good results on MoS$_2$-based systems.\cite{Gonzalez} We stress that the relaxed structure and electronics density of states obtained with \texttt{FIREBALL} are in good agreement with those obtained using the \texttt{VASP} code also used in this work.

\begin{suppinfo}
Supporting Information includes a discussion of the corrections of the optical spectroscopy data from optical interference effects, atomic force microscopy measurements, electron energy loss spectroscopy, power-dependent photoluminescence measurements, and DFT simulations of electronic density of states and STM images for various defects.

\end{suppinfo}

\section{Author Information}
*E-mail: johann.coraux@neel.cnrs.fr

\section{Associated Content}
The authors declare no competing financial interest.

\begin{acknowledgement}

This work was supported by the European Union H2020 Graphene Flagship program (grants no. 604391 and 696656) and the 2DTransformers project under OH-RISQUE program (ANR-14-OHRI-0004) and J2D (ANR-15-CE24-0017) and DIRACFORMAG (ANR-14-CE32-0003) projects of Agence Nationale de la Recherche (ANR). G.N. and V.B. thank support from CEFIPRA. The STEM (imaging and EELS) studies were conducted at the Laboratorio de Microscop\'{i}as Avanzadas, Instituto de Nanociencia de Arag\'{o}n, Universidad de Zaragoza, Spain. R.A. gratefully acknowledges the support from the Spanish Ministry of Economy and Competitiveness (MINECO) through project grant MAT2016-79776-P (AEI/FEDER, UE) and from the Government of Aragon and the European Social Fund under the project `Construyendo Europa desde Aragon' 2014-2020 (grant number E/26). We thank Jacek Kasprzak, Tomasz Jakubczyk, Maxime Richard and Le Si Dang for insightful discussions. C.G. acknowledges financial support from Spanish Ministry of Economy and Competitiveness, through the Mar\'{i}a de Maeztu Program for Units of Excellence in R\&D (Grant No. MDM-2014-0377).

\end{acknowledgement}

\newpage

%\bibliography{HP_MoS2}

\begin{mcitethebibliography}{81}
\bibitem[not()]{note_on_components}
The corresponding component is not the A$^-$ found in natural MoS$_2$. In fact,
  such a component is also expected (the main excitonic feature would hence
  consist of three components), but appears to represent a negligible one in
  case of HP/HT MoS$_2$.\relax
\mciteBstWouldAddEndPunctfalse
\mciteSetBstMidEndSepPunct{\mcitedefaultmidpunct}
{}{\mcitedefaultseppunct}\relax
\EndOfBibitem
\end{mcitethebibliography}

\providecommand{\latin}[1]{#1}
\providecommand*\mcitethebibliography{\thebibliography}
\csname @ifundefined\endcsname{endmcitethebibliography}
  {\let\endmcitethebibliography\endthebibliography}{}

\end{document}